\documentclass[useAMS,usenatbib]{mn2e}
\usepackage{graphicx}
\usepackage{epstopdf}
\title[kHz frequencies of accreting magnetospheres]
{On kHz oscillations and characteristic frequencies of accreting
magnetospheres}
\author[M. Ali Alpar]{M. Ali Alpar\thanks{E-mail:
alpar@sabanciuniv.edu}\\Sabanc{\i} University, Orhanl{\i}, Tuzla
34956, \.{I}stanbul, Turkey}

\begin{document}



\maketitle

\label{firstpage}

\begin{abstract}
When an accreting star is close to rotational equilibrium between
the dipole component of the stellar magnetic field and the accretion
disk, the star's rotation rate is roughly of the order of the Keplerean
rotation rate at the inner boundary of the disk, estimated as the
conventional Alfven radius. A range of frequencies higher than this
equilibrium rotation frequency can naturally arise if the accretion
flow is channeled by higher multipoles of the star's magnetic field.
The higher multipole components of the magnetic field will
balance the material stresses of the accretion flow at radii closer
to the star. The Kepler frequencies associated with these
generalized Alfven radii increase with the order of the multipole.
Other frequency bands, like the epicyclic frequencies associated
with the accretion flow, may in turn be higher than the Kepler
frequencies. We present expressions for the spectrum of higher
frequencies arising due to these effects. Kilohertz quasi-periodic
oscillation frequencies that are much higher than the rotation rate
of the neutron star, as observed from the recently
discovered 11 Hz (P = 90 ms) X-ray pulsar IGR J17480-2446 in the
globular cluster Terzan 5, may be due to modulation of the accretion
rate by the excitation of these modes in the accretion flow. The
very high QPO frequencies observed from the soft gamma repeaters SGR
1806-20 (P = 5.2 s) and SGR 1900+14 (P = 7.5 s) may also correspond
to these characteristic frequencies if SGRs accrete from fallback
disks around them. 
\end{abstract}


\section{Introduction}
Neutron stars accreting from disks tend towards a state of rotational equilibrium with the disk. 
In the equilibrium state the rotation rate of the star is equal or close to the Keplerean rotation rate at the inner boundary of the disk, 
which is the Alfven radius where the magnetic stresses of the ${\it dipole}$ magnetic field of the star balance the material stresses in the disk accretion flow and thereby disrupt the disk.
The recently discovered accreting and 
bursting pulsar IGR J17480-2446 in the globular cluster Terzan 5 attracts attention with its high
frequency quasi-periodic oscillations (QPO) at frequencies of 48, 173 and 814 Hz, much higher
than the rotation frequency of 11 Hz (Papitto et al 2011). The rotation rate itself
stands out as rather slow for neutron stars in globular clusters
which typically contain radio millisecond pulsars and accreting
X-ray millisecond pulsars. IGR J17480-2446 is still undergoing spin-up by accretion (Patruno et al 2011). 
The identification of the high QPO frequencies is an intriguing question.
A combination of very slow rotation rate and 
very high frequency QPO is also observed from the soft gamma repeaters SGR
1806-20 (P = 5.2 s ) and SGR 1900+14 (P = 7.5 s), which may be surrounded by an 
accretion flow according to the fallback disk model  (Alpar 2001; Chatterjee, Hernquist \& Narayan 2000). 
The purpose of this paper is to explore the higher frequency scales associated with the higher
multipoles in a magnetosphere in rotational equilibrium with the
disk at the dipole Alfven radius.
\section{Generalized Alfven Radii and Associated Frequencies}
The boundary between an accretion disk and the
magnetosphere of a star is primarily determined by balancing the
magnetic stresses due to the longest range, dipole, component of the
star's magnetic field against the material stresses of the accretion
flow in the disk, yielding the conventional Alfven radius
\begin{equation}
r_{A,1}= (\frac{{\mu_1}^2}{\phi_1 \dot{M}})^{2/7}{(GM)}^{-1/7}.
\end{equation}
Here the subscript refers to ${\it l=}$ 1 for dipole, $M$ is
the mass of the neutron star, $G$ the gravitational constant and
$\mu_1$ is the dipole magnetic moment of the star. Thus $r_{A,1}$ is
the stopping radius at which the dipole magnetic field of the star
disrupts the quasi-Keplerean flow of the material in the accretion
disk. The Alfven radius is defined by the balance of the material
stresses $\sim 1/2 \rho {v_{K}(r)}^2$ against the magnetic stresses
$\sim B(r)^2/8\pi$. The density $\rho$ arriving at the Alfven radius is
related to the mass transfer rate $\dot{M}$ arriving from the disk:
\begin{equation}
4\pi r_{A,1}^2 \rho v_{K}(r_{A,1}) \equiv \phi_1 \dot{M}
\end{equation}
where the factor $\phi_1$ describes the geometry of the
non-spherical flow, and also serves to scale the velocity of the
flow to the Kepler and free fall velocities, $v_K$ and $v_{ff} =
\sqrt{2} v_K$. (I have scaled the definition of $\phi_1$ with $v_K$
rather than $v_{ff}$ to incorporate the factor of $\sqrt{2}$.) 
Close to the dipole Alfven radius the accretion disk
has a boundary or transition region within which the rotation rate
$\Omega$ adjusts from Keplerean rotation towards corotation with the star
and magnetosphere. The term 'transition region' will denote a region, 
not necessarily extremely thin,
where the flow adapts to a new configuration. Inward of the dipole Alfven radius, 
at $ r < r_{A,1}$, the magnetosphere is loaded with accreting matter which follows 
magnetic field lines towards the star, acquiring some radial velocity commensurate with 
free-fall and gradually adopting to co-rotation, starting from quasi-Keplerean azimuthal velocities beyond $r_{A,1}$.

The actual magnetosphere-disk interaction and the behaviour of the accretion flow is likely to be
rather complicated. The magnetic axis of the neutron star is in general not aligned with the rotation axis. Moreover, the stellar field is likely to contain higher multipoles. 
There is evidence for the presence of higher multipoles of the magnetic field in classical T-Tauri stars 
(eg. Donati et al 2011a,b and references therein), and accreting 
white dwarfs (eg. Euchner et al. 2006, Beuermann et al. 2007).  Multipole magnetic fields on neutron stars are required by some models of 
pulsar emission mechanisms (Gil, Melikidze \& Mitra 2002; Mitra, Konar \& Bhattacharya 1999 and references therein), and in models for the pulse shapes 
of thermal X-ray emission from millisecond pulsars (Cheng \& Taam 2003). Mitra et al. (1999) have investigated the 
decay of neutron star magnetic fields by diffusion through the crust. They find that an initial field configuration with comparable strengths in the different multipoles 
decays and settles to a steady configuration at a few percent of the initial field strengths after decay timescales of 10$^4$-10$^5$ yrs. 
Multipoles up to ${\it l} = 8$ remain comparable to the dipole component in the final state of the surface magnetic field. 
A distinction between dipole and higher multipole components of magnetars, and the idea that magnetar fields reside in the higher multipoles 
has been proposed in the context of the fallback disk model, most recently for SGR 0418 with its observed upper limit on the dipole component 
of the field far below the magnetar range (Alpar, Ertan \& \c{C}al{\i}\c{s}kan 2011 and references therein).  
The accretion flow onto a neutron star with higher multipoles and non-aligned magnetic and rotation axes will have strong directional dependence. 

Aspects of the accreting magnetosphere problem have been investigated in 3-D numerical simulations for
dipole fields (Koldoba et al 2002, Romanova et al 2002, 2004), and for configurations involving
quadrupole (Long, Romanova \& Lovelace 2007, 2008) and octupole (Long, Romanova \& Lamb 2012)
fields. The numerical investigations show spatially localized signatures of various distinct  distance scales and complicated funnel flows. 
As in the example of simulations of accretion onto a magnetosphere with tilted dipole and quadrupole fields (Long, Romanova \& Lovelace  2008) 
these flows can be almost stationary. Unstable accretion has been investigated for accretion onto dipole fields 
(Kulkarni \& Romanova 2008, Romanova, Kulkarni  \& Lovelace 2008). Transient tongues of accretion protruding
into the magnetosphere with complicated patterns are found. If higher multipoles are present together with the dipole, 
the instabilities encountered by the accretion flow facing the dipole field will let the flow penetrate the magnetosphere 
to interact with the higher multipoles at distances closer to the star. Simulations of accretion onto dipole fields have 
produced QPO (Kulkarni \& Romanova 2009, Bachetti et al 2010). Determination of the quality of these QPOs 
to compare with the high Q of the observed QPO is not yet possible, due to the limited time span of the simulations. 
As yet QPO have not been reported from simulations of accretion onto higher multipole fields. 

One can estimate the distance and frequency scales that correspond to 
the interaction between multipole components of the star's magnetic field and the accretion flow. 
Our purpose here is to explore these frequency scales, without going into the discussion of 
how particular frequencies might be excited. 
The radial dependence of multipole component ${\it l}$ of the
magnetic field is:
\begin{equation}
B_{{\it l}}(r) \propto \frac{\mu_{{\it l}}}{ r^{({\it l} + 2)}}
\equiv B_{{\it l},*} (\frac{R_*}{r})^{({\it l} + 2)}
\end{equation}
where $\mu_{{\it l}}$ is the ${\it l}$th multipole moment,
$B_{{\it l},*}$ is a fiducial magnetic field value on 
the star's surface, and $R_*$ is the stellar radius (Jackson 1998). 
For simplicity I will label the multipole fields only with ${\it l}$, rather than $({\it l}, \;m)$. 
${\it l} = 1, 2, 3$ correspond to dipole, quadrupole and octupole fields respectively. 
Individual components of the magnetic field 
will have the $\theta$ and $\phi$ dependence characteristic of the multipole. 
Generalized Alfven radii for pure single multipole fields were discussed earlier by Lipunov (1978) and  Long, Romanova \& Lamb (2011).
Psaltis \& Chakrabarty (1999) gave expressions for the  Kepler frequencies associated with the generalized Alfven radii for single multipole fields. The situation is more complicated when the field contains several multipole components. 
As the accretion flow approaches the star along the magnetic field lines 
it encounters varying magnetic field stresses in different
directions. In general the different multipole axes are not aligned with each other, 
so that there is no axial symmetry, $m \neq 0$, with respect to the dipole axis.
The surface on which material and magnetic stresses balance
will have a complicated geometry, depending on the relative
strengths of the different magnetic multipole moments and the
inclination between the magnetic and rotation axes of the star.
As higher multipole components of the magnetic field dominate closer
to the star's surface, one can delineate a succession of
(generalized) Alfven stopping radii for the higher multipoles. 
These stopping radii are the characteristic distances
of the complicated matter-magnetic field interaction in the
magnetosphere. At the generalized Alfven radius for each successive
multipole, the magnetic stresses change slope as a function of the
distance $r$ from the star. This leads to pockets of density
accumulation. The generalized Alfven radii for multipoles ${\it l}$
satisfy
\begin{equation}
r_{A,{\it l}} = (\frac{\mu_{{\it l}}^4}{(\phi_{{\it l}} \dot{M})^2
(GM)})^{\frac{1}{4{\it l}+3}}
\end{equation}
where
\begin{equation}
4\pi r_{A,{\it l}}^2 \rho_{{\it l}} v_{K}(r_{A,{\it l}}) \equiv
\phi_{{\it l}} \dot{M}
\end{equation}
relates the local density $\rho_{{\it l}}$ in the pocket to the
total mass accretion rate $\dot{M}$ through the factor $\phi_{{\it l}}$.  

Equation (4) is a simple extension of the formula for generalized Alfven radii for an accretion flow 
facing a pure single multipole field 
(Lipunov 1978, Long Romanova \& Lamb 2011), differing  by the factor $\phi_{{\it l}}$. When a single multipole is present, this factor depends on the orientation of the multipole axis with respect to the disk. With several multipole components the accretion flow balances against the total field. 
How the total mass accretion rate $\dot{M}$ splits into various accretion funnels towards the different multipole stopping radii 
and what $\rho_{{\it l}}$ and $ v_{K}(r_{A,{\it l}})$ are in each location depend in a complicated way on the relative strengths of the 
different multipoles and on the angles between the various multipole symmetry axes. When several misaligned multipoles are present, 
Eqs. (4) and (5) do not determine the stopping radii for each multipole independently of the others, with some simple geometrical assumption 
for $\phi_{{\it l}}$ reflecting a thin or thick disk or spherical geometry. Rather, all multipoles, with their different misalignment angles, must be taken into account jointly and consistently 
- a task that can be realistically undertaken only with numerical simulations. Analytical models developed for accretion onto TTauri stars can provide guidelines to the problem (Gregory et al 2006, Gregory \& Donati 2011 and references therein). These models use current free magnetospheres. The actual situation or the outcome of a numerical calculation can be 
expressed in terms of the physical  channeling fractions $\phi_{{\it l}}$ in the simultaneous presence of the multipole fields in interaction with the accretion flow, 
which enter Eq. (4) in the combination $B_{{\it l},*}^{2}/\phi_{{\it l}}$. These channelling fractions are likely to have a sensitive dependence on the geometry, and it is not possible to separate them from the physically interesting surface fields in estimates using Eq. (4). Instead, I shall employ below a qualitative picture of the interaction of the mass accretion with the superposition of multipole fields.

Without any accretion flow, the distance scales $r_{\it l}$ where the multipole field strength matches the dipole 
\begin{equation}
B_{{\it l}}(r_{\it l})=B_1(r_{\it l})
\end{equation}
are determined by the surface fields alone, 
\begin{equation}
\frac{r_{\it l}}{R_*}=(\frac{B_{{\it l},*}}{B_{1,*}})^{\frac{1}{{\it l}-1}}.
\end{equation}
The radii where the strength of multipole ${\it l}$ matches that of the previous multipole ${\it l}$ -1 are 
\begin{equation}
\frac{r'_{\it l}}{R_*}=\frac{B_{{\it l},*}}{B_{{\it l}-1,*}}.
\end{equation}
Equations (7) and (8) are for multipole fields in vacuum, with no reference to any interaction with an accretion flow. The distance scales $r_{\it l}$ and $r'_{\it l}$ refer to the balance 
between different components of the field and constitute only a rough guide for the stopping radi where the various multipoles assume the prime role in balancing and channeling the accretion flow.

In the following we shall make an estimation of the successive generalized Alfven radii based on a qualitative picture of the balance between the accretion flow and a multicomponent 
magnetic field. Each multipole field has its "easy" directions at $\theta, \phi$ values where it has a significant radial component. The flow guided along the field lines in these directions 
will also be accelerated towards the star by gravity.
Accreting matter loaded onto the dipole field lines at the conventional dipole Alfven radius will flow towards the poles, in the easy direction of the 
dipole field. When the quadrupole axis is not aligned with the dipole axis, the easy dipole directions towards the poles do not coincide with the easy directions of the quadrupole field. 
Those components of the magnetic stress energy tensor to which the dipole field has negligible contribution at these directons of the flow, are dominated by the contributions of the quadrupole field. These stress components become important in balancing the material stresses and channelling the flow as matter approaches the 
star and the quadrupole field strength increases. Thus the accretion flow will eventually face a barrier at a quadrupole stopping radius, of the order 
of the quadrupole Alfven radius. The newly important components of the stress energy tensor, 
contributed by the quadrupole field, should be strong enough to balance the material stresses which are of the order of the material stresses (and therefore the balancing magnetic stresses) at the dipole Alfven radius. Note that the overall energy density in the dipole field here at the quadrupole stopping radius can be significantly higher than its value at the dipole Alfven radius, but it is the stresses in new directions, introduced by the quadrupole field that define the balance between the magnetosphere and the accretion flow.
The flow will now be diverted by the dominant quadrupole component of the magnetic field, to follow the field lines in non-radial directions until the magnetic field lines curve, towards the star following the easy directions of the quadrupole field,  which in turn do not coincide with easy directions of the octupole field. As the octupole component starts to dominate in the total magnetic field the flow is again diverted at the octupole stopping radius, following the octupole dominated field in non-radial directions until now the octupole field lines start curving towards the star, along the easy directions for the misaligned octupole field, and so on, if there are higher multipoles. Figure 1 shows a simple sketch illustrating these ideas for a magnetic field with 
misaligned dipole and quadrupole components. The presence of several different stopping distances and associated accretion pockets is clearly seen in figures depicting the 3-D simulations 
performed for combinations of dipole, quadrupole and octupole fields (see Figures 4 \& 5 in Long, Romanova \& Lovelace 2007;  Figures 5, 9 \& 15 in Long, Romanova \& Lovelace 2008; and Figures 14 \& 17  in Long, Romanova \& Lamb 2011). 

\begin{figure}
\centering
\includegraphics[width=0.5\textwidth]{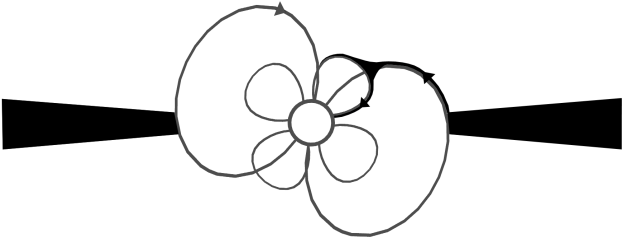}
\caption{Schematic depiction of the accretion flow from the disk onto the neutron star magnetosphere with misaligned dipole and quadrupole fields. The figure shows the plane of the rotation axis and the dipole symmetry axis. Only one path of accretion is shown. 
The symmetry is broken further when the quadrupole axis lies outside this plane. Dark lines indicate the locations of high density pockets and matter flow towards the star.}
\end{figure}

A surface of some complicated geometry will join the multipole stopping fronts with the conventional dipole Alfven surface. 
This convolved surface forms a boundary with the accretion flow with higher densities on the side away from the star.
The flow following the field lines forms pockets towards the star along easy directions of the multipole fields. (In the simple case of a pure dipole the easy directions lead to the polar caps). 
The generalized Alfven or stopping radii define the minima of this surface. At all
points on this surface the magnetic field balances the stresses from the accretion flow. The directions of the stresses change but we take the magnitude to be comparable to the pressure defining 
the largest stopping radius, the dipole Alfven radius at the maxima of the surface. This is still a rough estimate: the energy per particle will increase with decreasing radius due to fall in the gravitational potential well, while the density $\rho$ is likely to be less than the value at the inner edge of the accretion disk at the dipole Alfven radius, due to branching of the accretion flow as also represented in the unknown factors $\phi_{{\it l}}$. We take the relevant stress components over different parts of the 
surface to be comparable and of typical value given by $B^2/8\pi$, and postulate that the generalized Alfven radii for multipoles ${\it l}$ satisfy
\begin{equation}
\frac{B_{{\it l}}(r_{A,{\it
l}})^2}{8\pi}=\frac{B_1(r_{A,1})^2}{8\pi}
\end{equation}
which gives
\begin{eqnarray}
r_{A,{\it l}} & = & R_*
(\frac{B_{{\it l},*}}{B_{1,*}})^{2\xi/3}\;(\frac{r_{A,1}}{R_*})^{2\xi}\\
\frac{r_{A,{\it l}}}{r_{A,1}} & = & (\frac{B_{{\it
l},*}}{B_{1,*}})^{2\xi/3}\;(\frac{r_{A,1}}{R_* })^{-\eta}.
\end{eqnarray}
where
\begin{eqnarray}
\xi & \equiv & \frac{3}{2{\it l}+4} \\
\eta & \equiv & \frac{{\it l}-1}{{\it l}+2}.
\end{eqnarray}
In Equations (6) and (9) we have two different ways of estimating the multipole stopping radii. Equation (6) makes  a comparison of the multipole fıeld strengths as if thses were vacuum fields, without taking into account the presence of 
the accretion flow. Eq. (9) is based on balancing the magnetic stresses at the multipole stopping radii against the material stress on the convoluted boundary of the magnetosphere, as defined at the outermost, dipole stopping 
radius, the conventional Alfven radius. The results of both estimates were compared with the stopping radii for quadrupole plus dipole and octupole plus dipole fields 
as given in the figures depicting the simulations of  Long, Romanova \& Lovelace 2007, 2008 and  Long, Romanova \& Lamb 2011. The agreement with the simulation results is comparable for both models, 
of order 40\% for the quadrupole + dipole case and within a factor of two for the octupole + dipole case. This a satisfactory level of agreement, noting that we have neglected the angular dependence and 
dependence on relative orientation of the multipole axes. As the simulations do not indicate one model over the other we shall pursue the model taking into account the accretion flow. 

What are the characteristic frequencies in the accretion flow? The rotation rate of any rotating fluid plays a fundamental role in setting the frequency scales 
in dispersion relations. The rotation rate is keplerean in the far side of the accretion flow, and approaches the star's rotation rate as the flow approaches the star following closed corotating  field lines. 
The actual local rotation rate in the transition region will start with being keplerean and may remain commensurate with the local keplerean frequencies. Furthermore, as the matter loads onto magnetic 
field lines and falls toward the star, it will have velocities close to free-fall velocities, also of the order of the local keplerean velocities. The local Kepler frequency is thus the basic frequency scale in the 
accretion flow. This can be seen in the rotation rates obtained from simulations. For example Figure10 of Long, Romanova and Lovelace (2007), colour coded for rotation rates in dimensionless form, indicates 
rotation rates close to local Keplerean values in high density accretion pockets beyond stopping radii. The dispersion in free fall and Kepler velocities in these accretion pockets will lead to broad quasiperiodic 
frequency bands. To explain the quality of well defined QPO features the rotation rate $\Omega$ will be required to be modelled as it changes from Keplerian behaviour to co-rotation through the transition layer. 
The rotation rate in the accretion flow and related mode frequencies like the epicyclic frequency are expected to go through maxima as a function of position in the transition layer 
(Alpar \& Psaltis 2008, Erkut, Psaltis \& Alpar 2008). If these modes can be excited near the location that has maximum mode frequency, the shear $\partial \Omega/\partial r$ in the rotation 
rate and hence the dispersion in the mode frequencies will be minimal and well defined  QPO signals may be imprinted on the accretion flow.


The Kepler frequency at the conventional (dipole) Alfven radius
should be comparable to the star's rotation rate if the system is
close to rotational equilibrium. The Kepler frequencies at the
generalized Alfven radii for higher multipoles are higher than the
Kepler frequency at the dipole Alfven radius, with the ratios
\begin{eqnarray}
\frac{\Omega_{K}(r_{A,{\it l}})}{\Omega_{K}(r_{A,1})} & = &
(\frac{B_{{\it l},*}}{B_{1,*}})^{-\xi}\;
(\frac{r_{A,1}}{R_*})^{3\eta/2}
\nonumber \\
& = & (\frac{B_{{\it l},*}}{B_{1,*}})^{-\xi}\;
(\frac{\Omega_{K}(r_{A,1})}{\Omega_{crit}})^{-\eta},
\end{eqnarray}
where
\begin{equation}
\Omega_{crit} \equiv \frac{(GM)^{1/2}}{{R_*}^{3/2}}
\end{equation}
denotes the break-up frequency of the neutron star. The relative strengths of the multipole fields on the neutron star surface can be obtained by taking the star to be 
in rotational equilibrium with the disk at the dipole Alfven radius so that $\Omega_* = \Omega_{eq} \equiv  \Omega_{K}(r_{A,1})$ and the observed QPO 
frequencies are interpreted as  $\Omega_{K}(r_{A,{\it l}})$. Inverting Eq.(14) gives:
\begin{eqnarray}
\frac{B_{{\it l},*}}{B_{1,*}} & = & (\frac{\Omega_{crit}}{\Omega_{eq}})^{\eta/\xi}\; (\frac{\Omega_{eq}}{\Omega_{K}(r_{A,{\it l}})})^{1/\xi}
\nonumber \\
& = &  (\frac{\Omega_{crit}}{\Omega_{eq}})^{\frac{2{\it l}-2}{3}}\; (\frac{\Omega_{eq}}{\Omega_{K}(r_{A,{\it l}})})^{\frac{2{\it l}+4}{3}}
\end{eqnarray}
\\
The characteristic frequencies in accretion pockets above the
generalized Alfven radii can be much higher than the local Kepler
frequencies. As pointed out by Alpar \& Psaltis
(2008) for the boundary regions at the inner edge of an accretion
disk, the radial epicyclic frequency
\begin{equation}
\kappa = [2\Omega (2\Omega + r \frac{\partial\Omega}{\partial
r})]^{1/2}
\end{equation}
can be quiet different from the Kepler frequency. The rotation rate $\Omega$ 
deviates from the Keplerian rotation rate in a boundary or transition region 
between the disk and the magnetosphere corotating with the star. In the case of an accretion 
disk rotating faster than the neutron star and magnetosphere, $\Omega$ increases from 
$\Omega_*$ to $\Omega_{K}(r_{A,1})$ through the transition region at the dipole Alfven radius. 
The epicyclic frequency $\kappa$ is then greater than $\Omega$ and greater than even $\Omega_{K}(r_{A,1})$. 
Such sources  will be spinning up though they may be close to equilibrium, as expected for the low mass X-ray binaries. 
For sources rotating faster than the accretion disk at the dipole Alfven radius (sources in spin-down, like the SGR if these are acreting from fallback disks) 
Eq. (17) leads to epicyclic frequencies less than the Keplerian frequency at the dipole Alfven radius. However, 
the situation near the stopping radius for a higher multipole
component will always lead to epicyclic frequencies larger than the corresponding Kepler frequencies:  Matter is corotating with the star at the rate
$\Omega_*$ near the inner surface of the transition region while further out above the stopping radius the transition regions (accretion tongues or pockets) 
support velocities close to the local Keplerian values at $r_{A,{\it l}}$, which in turn 
are higher than the Keplerian velocities in the
disk beyond the conventional dipole Alfven radius $r_{A,1} > r_{A,{\it l}}$. Actual
frequencies of oscillation modes within an accretion pocket are
related to Kepler frequencies in a way that incorporates the various
magnetic or viscous forces, which may be represented by expressions
involving the rotation rate and its derivatives as well as sonic and
Alfvenic terms (Erkut,  Psaltis \& Alpar 2008). Between the stopping radius for multipole ${\it l}$
and the outer surface of the accretion pocket, where the rotation
rate is approximately Keplerian, there is a steep gradient of the
local rotation rate. From Eq. (17) we obtain the approximate
expression for the radial epicyclic frequency
\begin{eqnarray}
\kappa & \cong & \sqrt(2\frac{r}{\delta
r})\Omega \equiv f_{{\it l}}\;\Omega_K (r_{A,{\it l}})\\  
& = &  f_{{\it l}} (\frac{B_{{\it l},*}}{B_{1,0}})^{-\xi}
(\frac{\Omega_{*}}{\Omega_{crit}})^{-\eta}\Omega_{*}
\end{eqnarray}
where we have used $\Omega_{*} \cong \Omega_{eq} = \Omega_{K}(r_{A,1})$, for systems close to rotational equilibrium.

Similar arguments can be made for spherical accretion.
For spherical flows free-fall frequencies, which are of the same order as
Keplerian frequencies set the scale. The geometry of the magnetic
field breaks the symmetry, and forms accretion pockets above the
multipole stopping radii. The azimuthal and epicyclic frequencies
within the boundary regions should then scale as discussed above.
The generality of these arguments is reflected in the
factors $\phi_{{\it l}}$ in Eqs.(4) and (5).

\begin{table*}
\centering
\begin{minipage}{140mm}
\caption{High Frequency QPO from Slow Sources: Interpretation in terms of Multipole Magnetic Fields.
The column $\nu_{\it l} (Hz)$ lists the observed QPO frequencies which are associated here with modes at the stopping radii $r_{A,{\it l}}$. 
In the next column, the ratio of surface fields $\frac{B_{\it l,*}}{B_{1,*}}$ is given based on the assumption that the observed frequencies are Kepler frequencies. 
The next column gives the ratio $f_{\it l}$ of the epicyclic frequency to the Kepler frequency.  
For IGR J17480-2446, $f_{\it l}$ is calculated with the assumption that $\frac{B_{\it l,*}}{B_{1,*}} = 1$. For the SGRs $f_{\it l}$ is calculated with the assumption that  $\frac{B_{\it l,*}}{B_{1,*}} = 100.$}
\begin{tabular}{@{}rrrlrrrlr@{}}
\hline 
Source & $\,P (s)$ & $\,\nu_{*} (Hz)$ & {\it l} & $\,\nu_{\it l} (Hz)$  & $\,\frac{B_{\it l,*}}{B_{1,*}}\;(Kepler)$ & $\,f_{\it l}\;$ \\     
\\
\hline
IGR J17480-2446 & 0.09 & 11 & 2 & 48 & 0.60 & 1.2 \\
                             &         &      & 3 &173 & $\, 9.4\;10^{-2}$ & 2.0 \\
		       & 	        &      & 4 & 814 & $\, 9.3\;10^{-4}$ & 5.7 \\
\\
\hline
SGR 1860$-$20 & 5.2 & 0.19 & 2 & 18 & $\, 2.5\;10^{-3}$ & 53 \\
                          &       &          & 3 & 30 & $\, 9.9\;10^{-3}$ & 16 \\
		     &       &         & 4 & 92 &  $\, 1.7\;10^{-3}$ & 16 \\
                          &       &         & 5 & 150 & $\, 1.3\;10^{-3}$ & 11 \\
		    &       &         & 6 & 625 &  $\, 3.5\;10^{-6}$ & 25 \\
		    &       &         & 7 & 1840 &  $\, 1.1\;10^{-8}$ & 46 \\
\\
\hline
SGR 1900$+$14 & 7.5 & 0.13 & 2 & 28 & $\, 3.7\;10^{-4}$ & 109 \\
                           &       &         & 3 & 54 & $\, 6.7\;10^{-4}$ & 36 \\
		     & 	    &         & 4 & 84 & $\,1.2\;10^{-3}$ & 17 \\
		     & 	    &         & 5 &155 & $\,5.4\;10^{-4}$ & 13 \\
\hline
\end{tabular}
\end{minipage}
\end{table*}
\section{Discussion}
The multipole structure of a magnetosphere encountering an accretion flow can 
support a sequence of Keplerian frequencies  higher than the equilibrium rotation 
frequency of the star, as given in Eq. (14). For higher multipoles the frequency bands tend to
the Keplerian frequency on the stellar surface, which is the
break-up frequency of the neutron star, independently of the
relative strength of the high multipole fields and the actual
rotation rate of the star. A further effect towards producing high
frequencies arises due to the presence of epicyclic (radial) modes
in each of the accretion pockets above the generalized Alfven radii (Eqs. 18 \& 19).
These modes typically have frequencies higher than the quasi-Keplerian local azimuthal frequencies.

The Keplerian frequency spectrum corresponding to the various multipole stopping radii extends 
from $\Omega_{eq} = \Omega_{K}(r_{A,1}) \sim \Omega_{*}$ to the critical frequency $\Omega_{crit}$ which is of the order of $\sim 0.5\; ms$ for neutron
stars. The range of Keplerian
frequencies from $\Omega_{*}$ to $\Omega_{crit}$ can cover an expanded range
of several orders of magnitude for slowly rotating
stars, with a similar range between the
stellar radius and the size of the magnetosphere set by the
conventional (dipole) Alfven radius. If the surface magnetic field
in higher multipoles is large compared to the dipole, this also
serves for an extended range of frequencies and distance scales, as
the factor $(B_{{\it l},*}/\;B_{1,*})^{2\xi/3}$ in Eq. (10)
indicates. 

These higher frequencies are not detected from most slowly rotating accreting neutron stars. 
This may be because the imprint of higher frequencies on the accretion luminosity is too weak to be detected under
average accretion rates: Presumably it takes high accretion rates for enough
matter to be channeled through the higher multipole stopping radii
and the associated accretion pockets, so that the modes in these
pockets, carrying the high QPO frequencies, are excited to observable
amplitude only from  {\em bright} sources or during bright states of a source. Thus slowly rotating accreting neutron 
stars in high luminosity states are the prime candidates for displaying a wide range 
of well separated high QPO frequencies reflecting the multipole structure of the magnetosphere.
High mass X-ray binaries (HMXB), most of which are rather slow rotators,  have not exhibited high frequency QPOs. 
In most HMXB, the quasi-spherical accretion flow, captured from the stellar wind of 
the companion, produces a relatively weak X-ray luminosity 
which is noise-dominated, and may never be in a steady state or close to equilibrium.  
High frequency signals due to the higher multipoles may be too weak to be 
detected and may be dominated by timing noise. 

For the LMXB which harbor the fastest accreting neutron stars, the accreting X-ray millisecond 
pulsars, the range between the rotation frequency and the breakup frequency is 
rather small. Bands of QPO frequencies associated with the higher multipoles, if 
present, will blend in with the QPO features associated with the conventional dipole 
stopping radius and will not be distinctly identifiable.  

\subsection{ IGR J17480-2446, the 11 Hz pulsar in the globular cluster Terzan 5}

Among the LMXB the recently discovered source  IGR J17480-2446 in the globular cluster Terzan 5 with the slow rotation frequency of  11 Hz (Papitto et al 2011) would potentially have the
widest range of disjoint frequencies $\Omega_K (r_{A,\it{l}})$. 
This source has indeed displayed 
a range of QPO frequencies at 48 Hz, 173 Hz and 814 Hz. 

In Table 1 we display the inferences on the multipole field components derived by associating 
the observed QPO frequencies with Keplerian or epicyclic frequencies. 
The sequence of observed QPO frequencies is first evaluated with the simplest interpretation, 
associating the frequencies in increasing order with the Kepler frequencies at the stopping radii for ${\it l} = 2, 3, 4 ...$. 
The estimated values of the surface multipole field strengths in ratio to the dipole field strength, using Eq.(16), are given in the column labeled $B_{{\it l},*}/\;B_{1,*}(Kepler)$. 
The results show that for quadrupole and octupole components the surface fields are less than but within an order of magnitude of the surface dipole field.  The QPOs are  
likely to be associated with quasi-Keplerian frequencies against a surface field composition of comparable strengths in the multipoles.

Next, we assume that all multipoles have the same field strength on the neutron star surface, $B_{{\it l},*}/\;B_{1,*} = 1$, and calculate the factors  $f_{\it l}$ 
defined in Eq.(18) for epicyclic frequencies. We find that if the observed QPOs reflect epicyclic modes, the required transition regions are rather broad, 
with $r/\delta r \sim f_{\it l}^2 \sim 1.5 - 30$.

\subsection{Burst tail oscillations in the  SGRs 1806-20 and 1900+14}

The decaying tails following the giant flares of soft gamma
repeaters (SGRs; see reviews by Woods \& Thompson (2006) and
Mereghetti (2008)) are modulated at the star's rotation frequency,
revealing periods of 8 s, 7.5 s and 5.2 s for the SGRs 0526-66,
1900+14 and 1806-20 respectively (Mazets et al 1979, Hurley et al
1999, Palmer et al 2005). QPO at 18, 30, 92, 150, 625 and 1840 Hz were detected during the decay tail following the Dec 27 2004 giant flare 
of SGR 1806-20 (Israel et al. 2005, Strohmayer \& Watts 2006, Watts \& Strohmayer 2006).
Analysis of RXTE data on the decay tail of the 27 August 1998 giant flare of the SGR 1900+14  showed QPOs at 28 Hz, 54 Hz, 84 Hz 
and 155 Hz (Strohmayer \& Watts 2005). These QPOs have been
interpreted as modes of torsional oscillations of the neutron star
crust, characterized by the number of nodes ${\it n}$ in the radial
dependence of the shear wave, and the spherical harmonic indices
${\it l, m}$ up to ${\it l} =$ 11 (Watts \& Strohmayer (2007). 

SGRs and their cousins the anomalous X-ray pulsars (AXPs) could be accreting from a fallback disk (Alpar 2001, Ertan et al 2009 and references therein).  
In this model the neutron star is close to rotational
equilibrium with the disk, which explains the period clustering of
the SGRs and AXPs. The quiescent X-ray luminosity is due to
accretion while the soft gamma ray and X-ray bursts are taken to be due
to magnetar strength surface fields, following the standard magnetar model. The fallback
disk model makes a distinction between the surface magnetar fields, likely to contain strong higher multipoles, and the 
dipole field which governs the interaction with the disk and the
ensuing disk torques on the star. This dichotomy between the strengths of the dipole and higher 
multipole field components may reflect the creation, reconnection  
and amplification of the surface magnetic field in conjunction with crust breaking activity at scales 
small compared to the surface area of the neutron star. In an epoch of such activity the star could have higher multipole fields far in excess of the dipole field. 
The direct evidence for a dipole surface field strength far below the magnetar range comes from the SGR 0418+5729 (Rea et al. 2010), 
explained in terms of the fallback disk model by Alpar, Ertan \& \c{C}al{\i}\c{s}kan (2011).  Spectral modelling indicates a surface magnetic field 
in the magnetar range (G\"{u}ver, G\"{o}\u{g}\"{u}\c{s} \& \"{O}zel 2011), which must therefore reside in higher multipoles.

In terms of the fallback disk model, the post-burst accretion flows in  SGRs are candidates for the detection of high frequency QPOs from slow rotators in a state of high luminosity. 
During the burst tail, with still high luminosities partly due to accretion, the accretion flow disrupted by the burst would be re-adapting to the magnetospheric structure with its 
multipoles and characteristic frequencies. The X-ray bursters  testify to the possibility of accretion flows in the presence of high luminosity (super-Eddington) bursts. 
In Table 1 the high QPO frequencies observed from the SGRs 1806-20 and 1900+14 are evaluated as the Keplerian and epicyclic frequencies at the multipole Alfven radii in the accretion 
flow onto a  magnetar close to rotational equilibrium but rotating slower than the fallback disk at the conventional ${\it dipole}$ Alfven radius. 
The Kepler frequencies at the multipole stopping radii are substantially higher than the stellar rotation rate, and the epicyclic frequencies can therefore be 
even larger. The sequence of  frequencies is first evaluated with the simplest interpretation, 
associating the frequencies in increasing order with $\Omega_K (r_{A,\it l })$ for ${\it l} = 2, 3, 4 ...$. The ratios of the surface values for multipole fields to the surface 
dipole field, listed in the column labeled $B_{{\it l},*}/\;B_{1,*}(Kepler)$, are found to be $\sim$ 10$^{-3}$ or less, 
contrary to the expectation of the fallback disk model, and findings in the source SGR 0418+5729. If we now assume that multipole fields are 100 times as 
strong as the dipole field on the neutron star surface, and that the observed QPO frequencies correspond to epicyclic frequencies, the  
factors $f_{\it l}$ defined in Eq(18) turn out to have values in the 10-100 range, implying thin boundary layers, with $\delta r\;/\; r \sim 10^{-2} - 10^{-4}$.  
If SGRs are indeed accreting from fallback disks, the high frequency burst tail oscillations can be explained as epicyclic frequencies associated with multipole magnetar fields.
\section{Acknowledgments}
I thank the referee, Marina Romanova, for her careful review and helpful comments which improved the paper, and Onur Akbal and \c{S}irin \c{C}al{\i}\c{s}kan for help with the figure. 
The author of this paper is a member of the Science Academy, Istanbul, Turkey.

\end{document}